# The role of polymer architecture in the entropy driven segregation and spatial organization of bacterial chromosomes.

Debarshi Mitra [1], Shreerang Pande [1]& Apratim Chatterji [1]
[1] *Dept of Physics, IISER-Pune, Pune, India-411008.*
(Dated: December 22, 2021)

Entropic repulsion between DNA ring polymers under confinement is the key mechanism governing the spatial segregation of bacterial chromosomes, although it remains incompletely understood how proteins aid the process of entropic segregation. Here we establish that 'internal' loops within a modified-ring polymer architecture enhances entropic repulsion between two overlapping polymers confined in a cylinder. Moreover it also induces entropy-driven spatial organization of polymer segments as seen *in-vivo*. To that end, we design polymers of different architectures in our simulations, by introducing a minimal number of cross-links between particular monomers along the chain contour. This helps us to identify the underlying mechanisms which lead to faster segregation of spatially overlapping polymers as well as localization of specific polymer segments. The observed segregation dynamics of overlapping polymers is aided in our simulations by the frequent release of topological constraints, implemented by allowing chains to cross each at regular intervals. Additionally, we compare the segregation dynamics timescales with that of self-avoiding polymers and thereby highlight the significance of Topoisomerase in biological systems where DNA-strands are allowed to occasionally pass through each other. We use the blob model to provide a theoretical understanding of why or how certain architectures lead to enhanced entropic repulsive forces between loops, which in turn leads to the positional organization of segments relative to each other in confined environments. Lastly, we establish a correspondence between the *C.crescentus* bacterial species and our results for one particular polymer architecture.

The hitherto established paradigm of bacterial chromosomes being organized as amorphous polymers, has now been deemed obsolete[1]. In this vein, we have proposed previously that the *E.coli* chromosome can adopt particular polymer architectures aided by bridging proteins to enhance segregation dynamics and organize its genetic loci along the cell-long axis as seen in experiments [2]. Bacterial chromosomes have been typically considered as being ring polymers [3–6] and experimental evidence suggests that the bacteria *E.coli* plausibly uses cross-links created by SMC (structural maintenance proteins) to create an internal looped structure of the chromosome [7–9]. We show that this leads to the spatial localization of genomic loci as a consequence of the modified polymer architecture. It remains however to identify the pool of architectures which may be adopted by bacterial chromosomes which could drive faster segregation and also consequently leads to the localization of polymer segments as observed *in vivo*. Thus a detailed understanding of generic principles pertaining to the role of modified-ring polymer architectures in the entropy-driven localization of polymer segments is imperative to understand bacterial chromosome segregation at large. Such principles may also be employed to design synthetic polymers with tunable localization properties which have key technological ramifications.

It has been shown that the entropic repulsion between side-loops situated along the backbone of a bottle-brush polymer leads to a spontaneous emergence of helical order under confinement [10]. Though flexible polymers are considered amorphous with ability to adopt all possible conformations to maximize entropy, we show that adapting different polymer architectures with just 2 to 4 cross-links (CLs), which give rise to loops in a confined ring polymer can entropically lead to a particular internal organization of polymer segments. Consequently, there also occurs corresponding localization of polymer segments (genetic loci), with relatively small fluctuations in positions, along the cylinder long axis. Here we present a systematic analysis of some of the modified ring-polymer architectures and elucidate general principles concerning the organization and segregation dynamics of two overlapping polymers of various architectures under cylindrical confinement. We justify our computationally obtained results for different architectures, with an analytic treatment of the Helmholtz free energy difference between the overlapped and the segregated states and find that the "renormalized Flory approach" [11] successfully explains our computational observations.

In addition we also show that a subset of these polymer architectures could be potentially adopted by different bacterial species. The cylindrical confinement in that case represents the cell walls or the nucleoid within which the chromosome is confined within the cell.

To ensure efficient and timely segregation of initially overlapping polymer chains it is important that topological constraints are released. This is of utmost importance in the case of bacterial chromosomes where newly replicated chromosomes may be hindered from segregating due to topological constraints. The release of topological constraints in the case of bacterial chromosomes is accomplished by the presence of the enzyme DNA Topoisomerase (Topo II) which allows newly replicated daughter chromosomes to pass through each other akin to phantom chains [12–14]. Thus a comparative study investigating the dependence of topological constraint release on the dynamics of the segregation of initially overlapping polymers under confinement, is in order.

In the *E.coli* and *C.crescentus* bacterial cells, certain genomic segments of the chromosome are localised around certain positions along the cell long axis. Furthermore, detailed chromosomal contact maps obtained from Hi-C experiments [[15]] unambiguously have established the presence topological domains characteristic of different bacterial species. Such localisation patterns cannot be explained from entropic models of segregating ring polymers. The replication of chromosomes proceeds bidirectionally from the polymer segment (genomic loci) labelled as *oriC* and ends at *dif-ter*, to create two daughter DNA polymer chains. At the start of replication the *oriC* is at the mid-cell position [[7], [16]–[19]] in the case of *E.coli*. Post replication, the DNA-polymer organization changes so that the *dif-ter* locus moves to the middle with the two *oriC* loci occupying the 1/4-th and the 3/4-th positions along the cell longitudinal axis. The two *dif-ter* loci of the two daughter DNA-polymer remain cross-linked till the time of cell division, which occurs a few minutes after replication is complete (in the simpler slow growth conditions).

We showed previously that internal loops which form due to cross-links (created due to the bridging action of proteins *in-vivo*) between different monomers of the chain contour localises the *oriC*s at the quarter positions due to entropic repulsion between the internal loops[[2]] for *E.coli* chromosomes. Another previous study has shown that the presence of loops lead to a reduction in likelihood of chromosomes to get topologically entangled within the nucleoid/nucleus [[20], [21]]. The presence of macrodomains (spanning regions of. *Mbp*) within bacterial chromosomes has also been observed [[15], [22], [23]]. These may arise as a consequence of loops at specific positions in the genome [[2]]. These loops may be crucial to the formation of macrodomains and other topological features seen *in-vivo* [[2], [8], [24], [25]].

We also extend our studies on polymer architectures to the bacterial species of *C.crescentus*. To that end we calculate the contact map from our simulations with a particular polymer architecture, and identify that the contact map is strikingly reminiscent of the Hi-C contact map of the *C.crescentus* bacterium. Furthermore we find that the same architecture also results in the localisation of genomic loci with patterns very similar to those seen in other fluorescence based complementary experiments. Previous studies, which have tried to reproduce contact maps of bacterial DNA, optimise a large number of constraints, to find a detailed match with the experimental contact map[[26], [27]]. We take an orthogonal approach and focus more on the mechanism underlying the spatial localization of loci in our study, and try to match large scale macro-domain structures only, since a detailed match would require a more detailed simulation model which might not be particularly insightful to unearth general principles.

Some previous experiments on bacterial chromosomes as reported in [[2], [28], [29]], involved the tagging of certain loci with fluorescent markers, keeping the loci positions distributed over the entire ring contour. We plot the distribution of the monomers corresponding to the same loci, which were chosen experimentally. though we have access to the position distribution of all 500 monomers. Thereby, we can analyze the spatial distribution of any monomer (corresponding to a particular locus) that we need to, as we do when we compare contact maps or spatial distributions of loci, of *C.crescentus*. We also study how the release of topological constraints infrequently ( as is relevant biologically) aids the segregation process of the two (initially) overlapped polymers, and how the release of such topological constraints affects the organization of the polymers along the cylindrical long axis. Thus we deviate significantly from past theoretical studies in the polymer physics literature which usually incorporate excluded volume interactions between chains to prevent chain crossing.

In the remainder of this manuscript we first describe our simulation model in detail and then describe our findings in the 'Results' section. At the end of the 'Results' section we describe the key insights we extract from our *in-silico* studies. Subsequently in a separate section titled 'Relevance to *C.crescentus*' we establish the relevance of some of our results for the bacterial species *C.crescentus*. We end the manuscript with the 'Discussion' section where we summarise our work and mention avenues of further investigation.

### Simulation model

We perform Monte Carlo simulations using a bead spring model of a flexible ring polymer with 500 monomers in each chain. We introduce additional crosslinks to obtain polymers of different architectures. Two neighbouring monomers along the chain contour interact *via* the harmonic spring potential with energy $V_H = \kappa(r-a)^2$, where $r$ is the distance between the monomers. The unit of length in our study is $a$, where $a = 1$ is the equilibrium length of the harmonic-springs between two neighbouring monomers. The spring constant $\kappa$ is $100 k_B T/a^2$. The excluded volume (EV) interactions between monomers are modelled by the WCA (Weeks Chandler Anderson) potential and the diameter of each monomer is given by, $\sigma = 0.8a$. The form of the WCA potential is,

$$V_{WCA} = 4\epsilon[(\sigma/r)^{12} - (\sigma/r)^6], \forall r < 2^{1/6}\sigma. \quad (1)$$

Note that the $250th$ monomer of the two polymers are crosslinked to each other. This is inspired by the case of bacterial chromosomes where the newly replicated daughter chromosomes remain conjoined at the *dif-ter* locus prior to cell division, refer Fig.[1] [[2], [5]]. The polymers are confined to a cylinder with appropriately chosen dimensions. Furthermore, topological constraints are released frequently by decreasing bead diameter sizes. The organization of polymer segments is surmised from looking at the spatial probability distributions of a few chosen monomers along the chain contour. For more details about the simulation model, the reader is urged to refer to the 'Methods' section.

| Architecture | Pairs of crosslinked monomers | |
|---|---|---|
| Arc-0 | no additional CL | |
| Arc-2 | (125,1), (375,1). | |
| Arc-2-2 | (125,1) | (375,1) |
| | (156, 218) | (282,344) |
| Arc-3 | (125,1) | |
| Arc-4 | (200,50) | |
| Arc-6 | (150,230) | (350,270) |
| Arc-7 | (20,100) | (400,480) |
| Arc-8 | (100,1) | (125,200) |
| | (200,300) | (375,300) |
| Arc-9 | (125,1), | (125,375), |
| | (187,312) | |
| Arc-10 | (20,100), | (480,400), |
| | (187,313) | |
| Arc-11 | (125,1), (187,313) | |

Table I. Table listing the sites of crosslinks corresponding to different polymer architectures. The entries enclosed by the parentheses in the column on the right, represent the indices of the monomer pairs which are additionally crosslinked to obtain different polymer architectures as shown in Fig.2. The monomers labeled *oriC*, L1, L2, R1, R2, R3 and *dif-ter* in the ring-polymer are at positions 1, 22, 141, 443, 369, 304 and 250 along the chain contour.

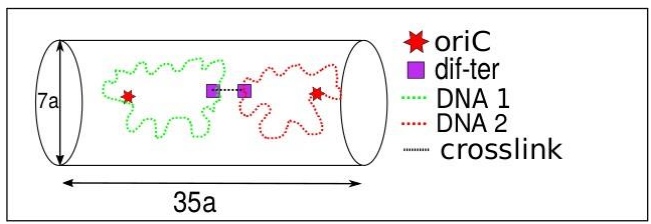

Figure 1. **A schematic representation of our simulation model** Two ring polymers (modified suitably with additional crosslinks) of 500 monomers are confined within a cylinder of aspect ratio 1 : 5 and thus a colloidal volume fraction of 0.2 is maintained. The two segments of each polymer chain labelled as *dif-ter* (following the convention for *E.coli* chromosomes) and represented by monomer index 250 (and 272 for the case of *C.crescentus*) , have been crosslinked . This is inspired by the fact that newly replicated chromosomes in bacterial cells remained conjoined at the 'ter' prior to cell division.

**Modifying the ring-polymer architecture** We now describe how we modify a ring polymer suitably to obtain different polymer architectures. We accomplish this by introducing additional crosslinks by introducing harmonic spring interactions between monomer pairs at specific sites along the chain contour. The spring constants corresponding to the additional crosslinks are given by $\kappa = 100 k_B T/a^2$. This results in a variety of looped structure of the polymers. We keep a minimal set of additional crosslinks and ensure that we do not introduce more than 4 additional crosslinks. In the table 1 we list the sites of crosslinks by which we generate different polymer architectures, which may be visualized better by referring to Fig.2. The polymer architectures under consideration represent a variety of looped structures of the modified ring polymers and the sizes of the loops vary from about 60 monomers to 130 monomers.

We start with just a ring polymer which we name Arc-0. Then we introduce a single crosslink (2 loops) placed at different sites along the contour such as Arc-3 and Arc-4. Arc-3 has a 'linear' looped architecture where the two loops are placed along the cylindrical axis. Arc-4 on the contrary has a 'side loop' architecture. This enables us to study the effect of loop orientation on the organization and segregation of spatially confined polymers.

In order to study the effect of additional loops we design architectures with 3 or 4 loops and these loops are oriented in different ways as before. The polymer architectures Arc-6, Arc-7 and Arc-8 have side loops of varying loop sizes. These side loops are located at different points along the chain contour. On the contrary polymer architectures like Arc-3, Arc-5 and Arc-11 have a looped structure where the loops are linearly arranged. The polymer architectures Arc-2, Arc-2-2, Arc-9 and Arc-10 may be considered as architectures having a combination of 'side-loops' and 'linear' loops.

**Importance of the 'ter' crosslink** Note that the 'side-looped' architectures can be distinguished from the 'linearly-looped' architectures, only because the two initially overlapping polymers are conjoined at the 'ter', which restricts the number of conformations accessible to the two polymers. For instance Arc-3 and Arc-4 are not equivalent only due to the presence of the *ter* crosslink.

### Simulation Results

**Dynamics of Segregation** We first present a comparison of segregation times of spatially overlapping polymers, for different polymer architectures. To estimate the segregation times we measure the distance between the center of masses of the two polymers from $t = 0$. If the mean of the distance between the center of masses fluctuates is greater or equal to a designated threshold, we deem the polymers to have segregated. The polymers are ideally deemed to have completely segregated if the distance between their center of masses is greater or equal to $0.5L$ where $L$ denotes the length of the confining cylinder. Thus in Fig.3a we show the mean segregation times for different architectures, without allowing chain crossing. On the right y-axis we show the success rate of segregation which indicates what percentage of independent runs result in segregation. We note from Fig.3a that architectures with additional cross-links lead to more efficient segregation in lesser times as compared to just a ring polymer (Arc-0). For instance in Fig.3a Arc-8 shows close to 60% success rate and low segregation times. Note that Arc-8 has 4 additional crosslinks. However interestingly Arc-5 and Arc-2-2 also shows a high success rate of segregation, but Arc-5 only has two additional crosslinks.

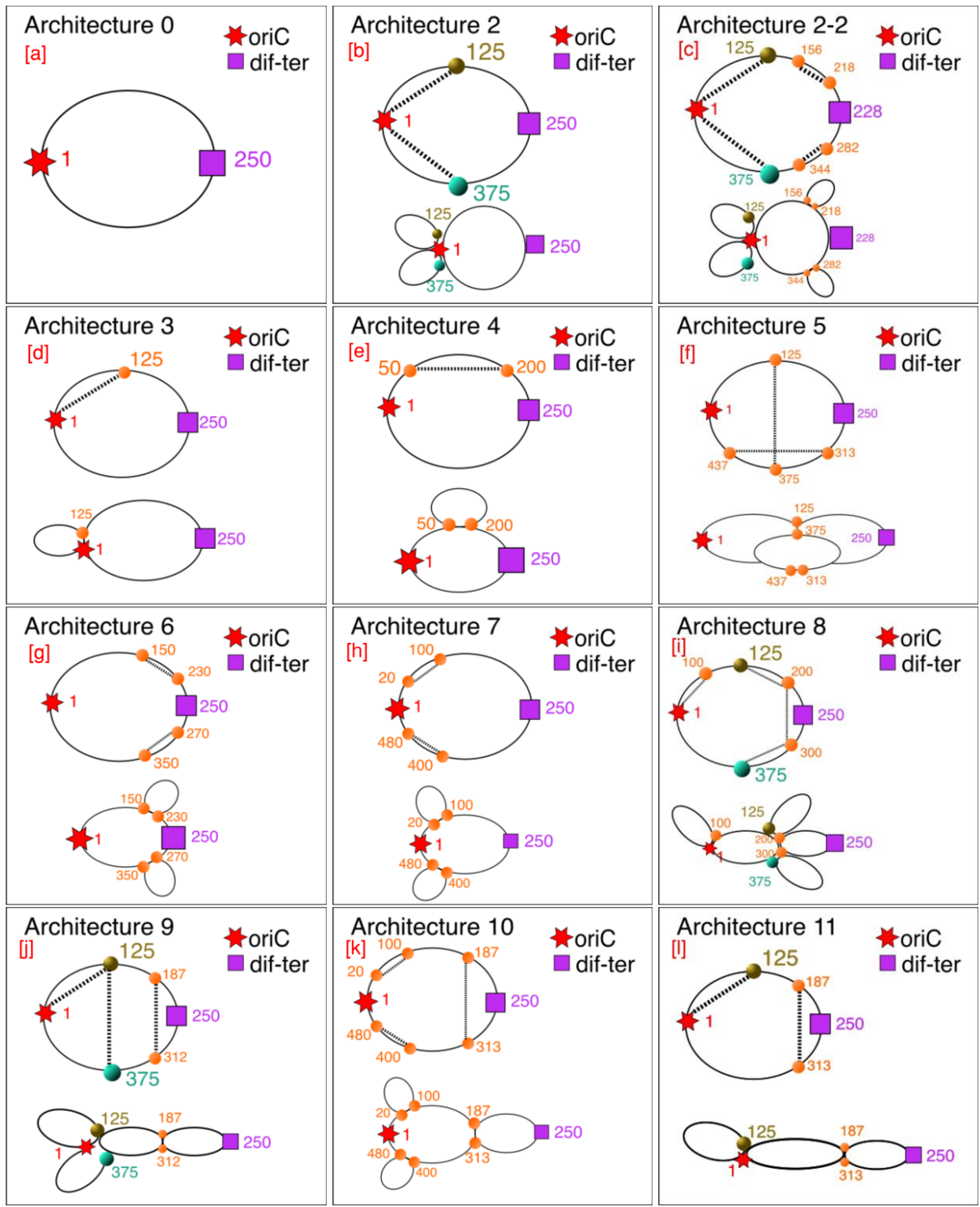

Figure 2. **Schematic representation of polymer architectures under consideration** (a) Arc-0, (b) Arc-2, (c)Arc2-2, (d)Arc-3, (e)Arc-4, (f)Arc5 (g)Arc-6, (h)Arc-7, (i)Arc-8 (j)Arc-9 (k)Arc-10 and (l)Arc-11

The sites of crosslinking in Arc-5 and Arc-2-2 thus lead to efficient segregation with a high success rate. This implies that certain architectures are more amenable to segregation of overlapping polymers by entropic means.

**Effect of topological constraint release** We also observe that release of topological constraints leads to increased success rates of segregation and lowers the times required for spatial segregation. For the data presented in Fig.3a, we do not allow the release of topological constraints by decreasing the sizes of the monomers, while for the data presented in Fig.3b we allow chain crossing. Additional details of these calculations have been provided in Sec-1, of the Supplementary (S.I). Furthermore the segregation times and success rates for a less stringent criterion have been provided in Sec-1, Fig.S2 of S.I. A comparison between Fig.3a and Fig.3b shows that the release of topological constraints is crucial to achieve higher success rates of segregation. We also note that even for the architectures that successfully show segregation, the time of segregation is much lower when topological constraints are released. Thus the release of topological constraints is pivotal to accomplish timely and efficient segregation of spatially overlapping polymers.

**Quantifying the propensity of polymers to remain segregated.** After the initially overlapping polymers 'demix' due to entropic conditions, could some architectures again get spatially overlapped to some degree due to thermal fluctuations? To investigate the same we look at the probability distributions of the distance between the center of masses of the two polymers $D_{com}$, for all the architectures, refer Fig.3c and Fig.3d. We note that Arc-0 (ring polymer) shows a relatively broad distribution compared to other architectures further indicating that it leads to inefficient segregation of daughter chromosomes. We also note that Arc-8 with 3 'side loops' shows the highest propensity to avoid spatial overlap.

**Spatial organization of polymer segments (genomic loci)** We now show that certain architectures lead to a specific organization of the segregated polymers along the cylindrical axis. We do this by looking at the spatial probability distribution of a few chosen monomers along the chain contour, refer Model section. In Fig.4 we show the distribution of the *oriC* and *dif* locus. To plot these distributions we compute the average distribution from the successfully segregated cases having incorporated the effect of Topoisomerase (which leads to release of topological constraints). Note that we show the distribution of only one of the *dif-ter* locus since the two *dif-ter* loci are crosslinked and therefore will have nearly overlapping distributions. In the Supplementary (refer Sec-4, Fig.S3 and S4) we show similar data but without allowing chain crossing. We note that the effect of chain crossing doesn't significantly affect the localisation of *oriC* and the *dif-ter* locus. Furthermore the spatial probability distributions of the other monomers (loci) are given in the supplementary(S.I) in Figs. S5, S6, S7 and S8. The mechanism by which different monomers (genomic loci) get localized due to the presence of loops, has been already discussed at length in [2] for architecture Arc-2. Here we show that not all looped architectures lead to the localisation of polymer segments and only a 'linearly' looped architecture leads to the localisation of polymer segments. In a polymer architecture having such 'linear' loops, the constituent loops repel each other entropically and thus the polymer segments (genomic loci) get localised to certain positions along the cylindrical axis.

We once again show that the effect of Topoisomerase which allows chain crossing, is inconsequential with regards to the localisation of the chosen monomers (genomic loci), refer Figs. S9, S10, S11 and S12, where we show spatial probability distributions of the chosen monomers without allowing chain crossing.

**Free energy analysis: The "renormalized Flory approach"** We now establish that our computational results are in agreement with an analytical treatment of free energy difference between the segregated and unsegregated states using a "renormalized Flory approach" [30–36]. Hence, we calculate the Helmholtz free energy difference per monomer $\delta F/N$ , between the overlapped and the segregated states of the two polymers, for different architectures, refer Fig.5. We show that certain architectures have a larger $\delta F/N$ and thus lead to an enhancement of entropic repulsive forces. Our analytical treatment considers polymers confined in an open cylinder, and is applicable only in the blob scaling regime where the diameter of the confining (open) cylinder, D, is greater than $10\sigma$ [34]. To incorporate the effect of different architectures in free energy calculations, we treat each loop (of each polymer) to be composed of two linear segments (comprising of half the number of monomers in the loop) trapped in tubes of a suitable "effective" diameter $D'$. The value of $D'$ is dependent on the architecture. We provide the details of these calculations in the supplementary, refer Sec-5. Note that the free energy difference between the overlapped and the segregated states of the polymers is proportional to $f_{bl}$, which is the free energy associated with each blob, and is inversely proportional to $D'^{5/3}$. In Fig.5 we plot the analytically estimated $\delta F/N$ corresponding to different architectures, for an arbitrary choice of $D = 10\sigma$ and $f_{bl} = k_B T$.

We note that Arc-3 is the least favourable when it comes to segregation of polymers and the $\delta F/N$ for segregation is about $3 \times 10^{-2} k_B T$ . Surprisingly, the value is even lesser than that of Arc-0. Thus the mere presence of cross-link(s) may not be sufficient to induce strong entropic forces of segregation. The sites of these cross-links therefore play a crucial role. We note that Arc-8 gives rise to the highest value of $\delta F/N$ among the investigated architectures. However as we noted earlier Arc-8 does not lead to the localisation of the *oriCs* to the quarter positions. Thus, the architecture adopted by the *E.coli* chromosome may not arise solely due to considerations of quick segregation and maybe influenced by other biological processes such as gene expression [27].

Note that our analytical free energy calculations is consistent to a fair degree, with previously presented simulation results in Fig.3. In Fig.3 Arc-8 shows the highest success rate and the least time required for segregation. This is consistent with our finding that the (for overlapped and segregated states) corresponding to Arc-8 is the highest. Similarly, Arc-3 and Arc-0 show relatively lower success rates and comparatively higher times for segregation. This too is consistent with our free energy results presented in Fig.5, although the results do not justify such a large difference in of segregation, between Arc-3 and Arc-0. We attribute this to the finite size effects in our simulations arising from a closed cylinder and relatively few monomers, for which the blob scaling picture might not be entirely accurate. Despite these caveats, our analysis is quite consistent with several of our previous findings presented before. Thus we believe that our free energy analysis, despite its limitations, is fairly useful with regards to understanding which polymer architectures are more amenable to efficient spatial segregation of spatially overlapping polymers under confinement.

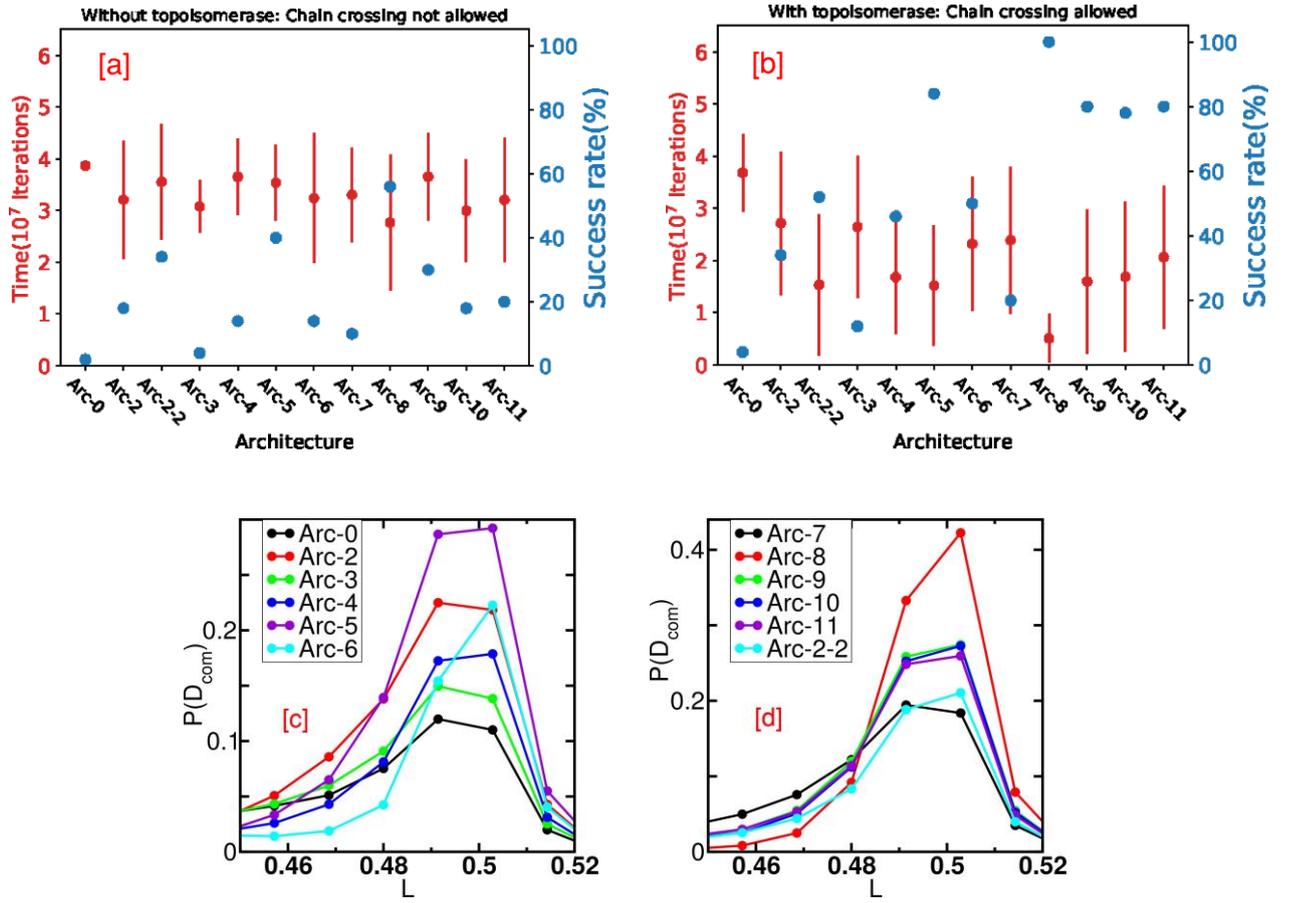

**Figure 3.** **Dynamics of segregation of overlapping polymers for different architectures** Fig. shows a comparison of segregation speeds for different architectures. The right y-axis shows the success rate of segregation which indicates what percentage of independent runs lead to segregation. The designated threshold criterion for considering two polymer chains to have spatially segregated is the average (over $4 \times 10^7$ MCS) of the distance between the center of masses, $< D_{com} >$ to have a value, greater or equal to $0.5L$, where $L$ is the length of the confining cylinder. For results with different criteria refer Fig.S2 in S.I. We note that the architectures with additional cross-links lead to more efficient and quicker segregation as compared to just a ring polymer (Arc-0).The shown data has been averaged over 50 independent runs. The error bars have been computed from the successfully segregated cases from the entire set of 50 runs. The effect of Topoisomerase (to release topological constraints) has not been incorporated in (a) while release of topological constraints is allowed in (b). In (c) and (d) we show the probability distribution of the distance between center of masses ($D_{com}$), computed from 50 independent runs (including the cases where segregation is unsuccessful). The x-axis has been scaled by the length of the confining cylinder (L). The probability distributions indicate which architectures lead to larger fluctuations in $D_{com}$ or which architectures have a larger propensity to mix due to fluctuations, even after being demixed due to entropy. The effect of topoisomerase (or chain crossing) has been incorporated for the data presented in (c) and (d).

**Principles of segregation and localisation of polymer segments due to polymer architectures** Thus we show that polymer architectures with additional crosslinks which suitably modifies the ring polymer architecture leads to an enhancement of entropic repulsive forces of segregation, and consequently lower segregation times. Through this manuscript we have also studied a variety of polymer architectures and established that some successfully organize polymer segments while some others lead to faster dynamics of segregation of initially overlapping polymers. Interestingly, we note that all the architectures with 'linear' loops (akin to linearly placed blobs in a cylinder) lead to a well-defined organization of the genomic segments. This may be confirmed from Fig.4 and Sec.3 in the S.I, where we find for instance that Arc-3, localizes genomic segments much better than Arc-4. Note that both Arc-3 and Arc-4 have one additional crosslink. Similarly Arc-6 and Arc-7 performs much worse with regards to localisation in comparison to Arc-11. Note that Arc-6, Arc-7 and Arc-11 have two additional crosslinks each. From Sec.3 in S.I we find that Arc-5 which is another 'linearly looped' architecture also efficiently localises polymer segments quite effectively.

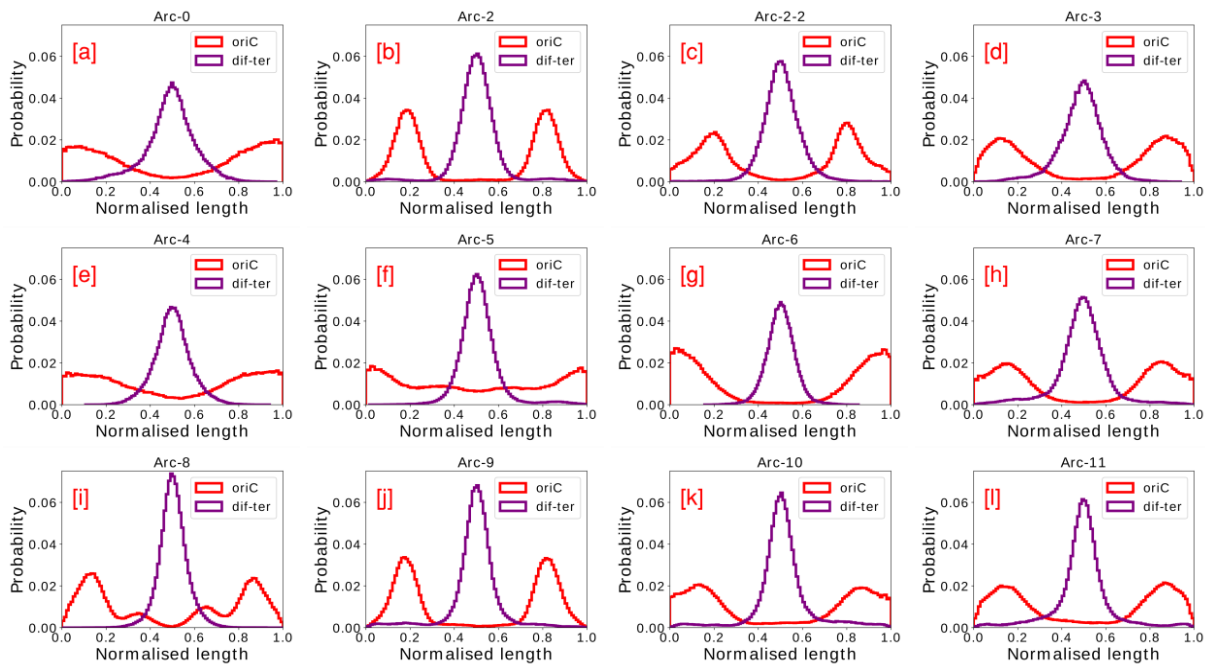

Figure 4. **Localization of chosen monomers (genomic loci) for all architectures** Fig. shows the distribution of the *oriC* positions for (a)Arc-0, (b)Arc-2, (c)Arc-2-2, (d)Arc-3, (e)Arc-4, (f)Arc-5, (g)Arc-6, (h)Arc-7, (i)Arc-8, (j)Arc-9, (k)Arc-10 and (l)Arc-11. We note that Arc-2, Arc-2-2 and Arc-9 shows a sharply peaked distribution of the *oriC* around the quarter positions and a sharply peaked distribution of the *dif* locus around the mid-cell position. The data has been produced by averaging over the runs that successfully lead to segregation, from a pool of 50 independent runs. Note that we show the distribution of only one of the *dif* locus since the two *dif* loci are crosslinked and therefore are expected to have nearly overlapping distributions. We allow chain crossing at infrequent intervals.

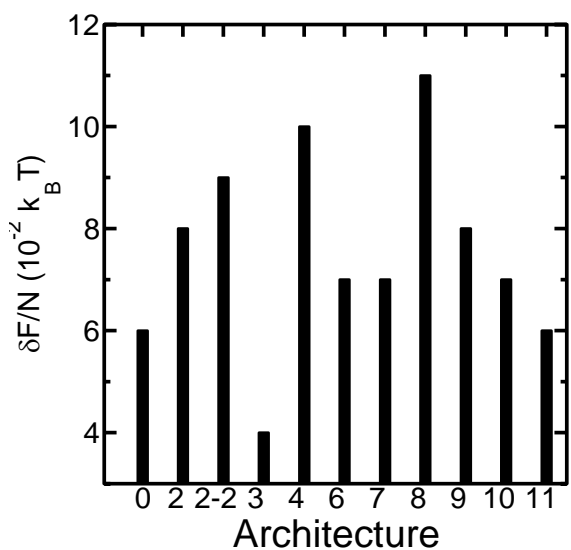

Figure 5. **Free energy calculations from the "renormalized Flory approach"** Fig. shows the analytically estimated Helmholtz free energy difference per monomer, $\delta F/N$ in units of $k_B T$ between the overlapped and segregated states of the two polymers, for several polymer architectures. We assume that the free energy associated with each blob is $k_B T$ and the diameter of the confining (open) cylinder is 10a.

The architectures which employ 'side loops' however lead to faster and efficient segregation dynamics. This may be surmised from Fig.3 where we find that Arc-4 shows a higher success rate as well as lower time of segregation, as compared to that of Arc-3, despite both these architectures having one additional crosslink each.

The architectures Arc-9 and Arc-2-2 which may be considered as hybrid architectures, having a combination of both 'linear' and 'side loops' however localise extremely efficiently and also lead to efficient spatial segregation. Such 'hybrid' loop structures thus are able to induce spatial segregation in relatively smaller times and also localise genomic loci efficiently. We have already established that Arc-2-2 is plausibly the polymer-architecture adopted by *E.coli* chromosomes [2]. In the subsequent subsection we employ Arc-9 to show that it is able to successfully reconcile results pertaining to *C.crescentus* chromosomes.

### Relevance to *C.crescentus*

Having discussed at length the role of polymer architecture in the organization and segregation of spatially overlapping polymers, we now wish to investigate whether a certain architecture namely Arc-9 is possibly adopted by *C.crescentus*. To that end we look at the spatial localisation patterns of loci measured in experiments albeit with the monomer corresponding to the *ori* locus tethered to the cell-pole, as is the case for *C.crescentus* chromosomes [1, 29, 37].

We present in Fig.6 (a) and (b) the spatial probability distributions of certain loci, obtained from experiments [29] and our simulations respectively. We find that we reproduce the localisation patterns seen *in-vivo* to a fair degree, despite the simplicity of our model. A previous study has already established that such localisation patterns cannot be obtained from just a confined ring polymer with its *ori* tethered to the cell-pole [1].

Furthermore *via* Fig.6(c) and (d) we show the contact maps obtained from experiments and simulations respectively. We find that our simulated contact map is in good agreeemnt with that obtained from simulations. Moreover we recover the other less prominent diagonal which is characteristic of *C.crescentus* Hi-C data[37].

For bacterial chromosomes such as *C.crescentus* it is most likely that there exist many other 'effective' crosslinks which may arise dynamically due to several proteins found *in-vivo* and plectoneme super-coiling [1]. However we propose that the crosslinks we have introduced are relatively long-lived and is responsible for the organisation of the *C.crescentus* chromosome.

The mechanism by which a particular polymer architecture loci local ised polymer segments (or genomic loci) has been already proposed by us in [2] for *E.coli* chromosomes. To put it simply, a 'looped' polymer architecture induces 'intra-polymer' entropic repulsion between loops of each polymer (or same sister chromosome), as well as 'inter-polymer' entropic repulsion between loops belonging to different polymers (or different sister chromosomes). This mechanism was established and examined thoroughly for different polymer architectures in this manuscript

**Discussions:** We establish the mechanisms by which polymer segments of a polymer get localised in a confined geometry by analyzing a pool of polymer architectures. Thus we show that by introducing strategically placed loops within a ring polymer, one can induce an "effective" repulsive interaction between particular polymer segments, which leads to specific spatial localization patterns of the same. Such entropy-driven "effective" interactions may find a host of applications where one has to design polymer systems with specific organizational properties, without invoking explicit enthalpic interactions. We also establish the principle by which modified polymer architectures lead to an enhancement of the entropic repulsion between overlapping polymers, and thereby aid the segregation process. We show this by estimating the free energy difference between overlapped and segregated states analytically for several polymer architectures using the "renormalized Flory approach" [84].

These modified architectures may arise in bacterial chromosomes to accomplish biological functions [27], but may also lead to the enhancement of the entropic forces of segregation. We also show that the release of topological constraints plays a crucial role in the speeding up of the segregation process and also results in a significant increase in the success rates of segregation. Furthermore several experimental results have shown previously that chromosomal loci show specific localisation patterns in bacterial chromosomes [18, 28]. Thus we show here that such localisation patterns observed for *C.crescentus* chromosomes, are consistent with those obtained with polymer architecture Arc-9. Previously, we established that architecture Arc-2-2 reconciles spatio-temporal localization patterns seen in experiments for the *E.coli* bacterial chromosome [2]. Thereby, this current manuscript further strengthens the thesis that bacterial chromosomes employ this generic principle of entropy based organization of polymeric segments, to organize itself within the cell.

Pelletier et al [6] had experimentally established previously the prevalence of crosslinks in bacterial chromosomes. These crosslinks may diffusively form and get disassociated, however we hypothesize that certain crosslinks at specific sites may be long lived, which affects the segregation dynamics of daughter chromosomes. We note from Fig.6c and (d) that the simulated contact map shows illuminated square patches which are absent from the experimental contact map. Previous studies indicate that such illuminated square-like patches would not appear, if within such 'loop domains' there are dynamic cross-links which get induced at regular intervals [24, 27]. Such dynamic cross-links will additionally induce entropic organization within such domains and may therefore lead to a closer resemblance of the simulated contact map to the experimentally obtained Hi-C contact map.

Future advances in nanotechnology may rely on the principles outlined here, of strategically designing polymer architectures to induce localisation of certain polymer segments. This will plausibly be of considerable technological importance especially for applications which require spatial localization of pharmaceutical cargo in confined environments [38, 39]. More fundamentally the approach of tuning entropic interactions between polymer segments *via* polymer architectures could also be em-

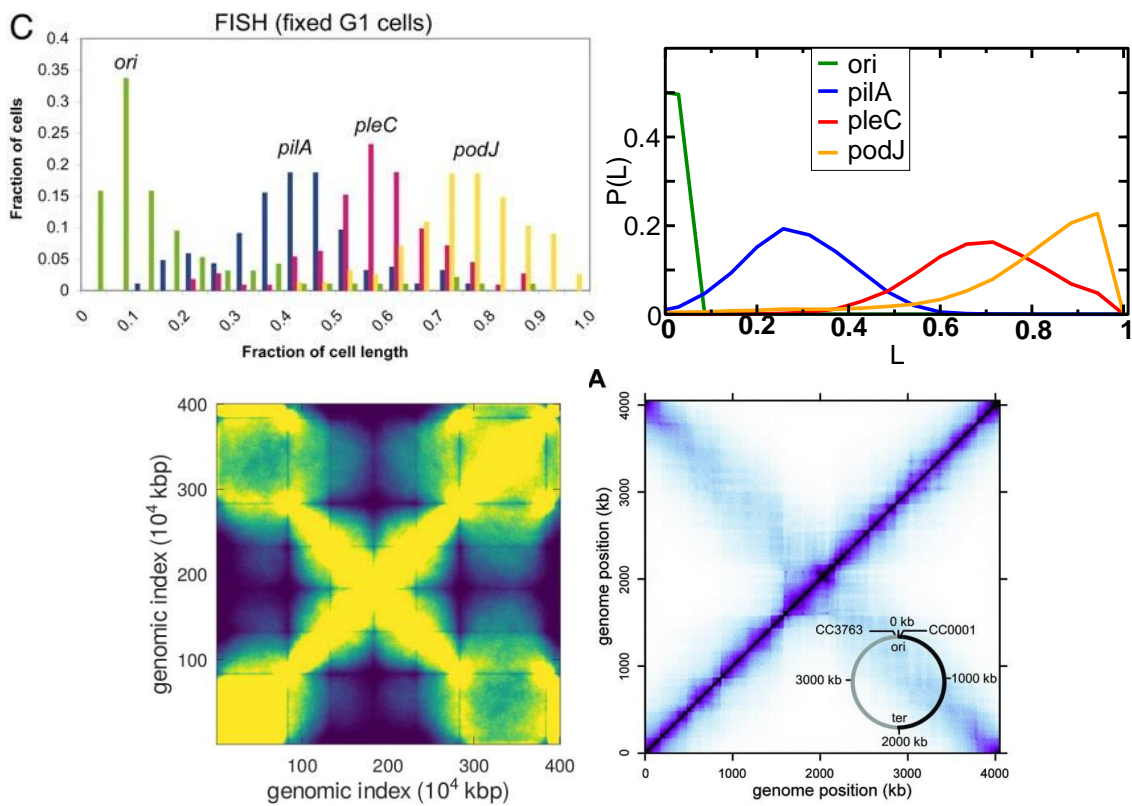

Figure 6. **Spatial organization of *C.crescentus* chromsome, investigated *via* simulations** (a) and (b) shows the spatial probability distributions of loci obtained from experiments and simulations respectively. In the simulations we use the polymer architecture Arc-9 with the monomer corresponding to the *ori* tethered to the cell pole, as is the case *in-vivo*. The x-axis has been scaled by the length of the confining cylinder. We do not show the distribution for the *ori* locus since it remains tethered to the cell-pole and therefore trivially remains localised around the pole [1, 29, 37]. Fig.6a has been reproduced here from [29], after having obtained due permission to do so.

Furthermore we also obtain the contact map from simulations (c) which matches the experimental Hi-C data shown in (d), to a great degree, including the less prominent diagonal [37]. Subfigures (a) and (d) have been adopted from [29] and [37] after having obtained requisite permissions.

ployed by researchers from diverse backgrounds to obtain micro-phase separation of polymeric mixtures induced solely due to entropic considerations without having to invoke enthalpic interactions between the constituent polymers.

## Methods

**Cylindrical confinement: Relevance to bacterial chromosomes** We confine two polymer chains within a cylinder of radius $3.5a$ and a length of $35a$, refer Fig.1. This ensures an aspect ratio of $1:5$, which is roughly what is observed in-vivo for an *E.coli* cell in slow growth conditions [40] with two daughter chromosomes, at the end of the replication process. This particular choice of parameters results in a colloidal volume fraction of monomers to be 0.2. If the polymer chains represent replicated *E.coli* daughter chromosomes and since the

*E.coli* chromosome has $4.6Mbp$ of DNA, each monomer corresponds to $9.2kbp$ of DNA. Note that the Kuhn length of bare ds-DNA is known to be $100nm$ with approximately 300 base pairs (bp). Thus each monomer corresponds to a length scale much greater than the Kuhn length. We do not consider larger cylinders as it is established that segregation occurs only when the radius of the confining cylinder is less than the radius of gyration $R_g$ of the polymer [5].

**Release of Topological constraints** We also consider polymer chains which can pass through each other at regular intervals, to release topological constraints. This is done to mimic the effect of Topoismerase in the *E.coli* cell which allows release of topological constraints by facilitating chain crossing. To implement this, we reduce the $\sigma$ to $\sigma' = 0.2a$ after every 10000 Monte Carlo steps (MCS). We let the system evolve with $\sigma = 0.2a$ for the subsequent 900 MCS, to allow the release of topological constraints, before reverting to $\sigma = 0.8a$.

**Quantifying the organisation of polymer segments (genomic loci)** To establish the organization of the polymers along the cylindrical long axis we plot the probability distributions of certain monomers along the cylindrical long axis. The chosen monomers are fairly equally spaced along the chain contour and hence reliably indicate the degree of (linear) organization along the cylindrical axis. We study the spatial localization of monomers: 1, 22, 141, 443, 369, 304 and 250 in each of the two polymer chains. If the two polymer chains are thought of as being analogous to newly replicated chromosomes in *E.coli* cells, then the position of these monomers along the contour correspond to loci labelled as *oriC*, L1, L2, R1, R2, R3 and *dif-ter* in [2, 28], specifically for the *E.coli* chromosome.

**Initialising and Equilibrating the system** As polymers spontaneously segregate due to EV interactions, we equilibrate the system of two ring polymers in an initial mixed state by the following procedure. We introduce the monomers at random positions in the cylinder and allow them to diffuse freely without excluded volume interactions between them. However the harmonic spring interactions are present (including the additional crosslinks). We run the simulation for $10^6$ MCS to equilibrate. Once the chains get completely 'mixed' spatially we switch on the excluded volume interactions and study the segregation dynamics of the two polymers. Note that here we do not model replication explicitly unlike in [2] and the dimensions of the confining cylinder is kept constant throughout the length of the simulation run.

**Modelling the *C.crescentus* bacterial species** We find that the confined polymer with architecture Arc-9 successfully organises the polymer segments in a way that is particularly relevant for *C.crescentus*. However the *ori* is tethered at the cell-pole for the case of the *C.crescentus* chromosome. In our simulation model for *C.crescentus* the monomer index 22 corresponds to the *ori*, which is tethered to the cylinder (cell) pole, using harmonic spring interaction with a fixed point at the pole. We also look at the spatial distributions of other loci such as *pilA*, *pleC*, *podJ* and *ter* [1]. These loci correspond to the monomer indices 122, 182, 238 and 272 respectively . We also look at the contact map obtained from our simulation model using the procedure outlined in [2]. We initialize the simulations with an equilibrated conformation of two polymers having the Arc-9 architecture. As mentioed before the *ori* is tethered to the cell (cylinder) pole *via* a spring harmonic interaction, with a spring constant $10k_BT/a^2$.

## AUTHOR INFORMATION

### Corresponding Author

**E-mail:** debarshi.mitra@students.iiserpune.ac.in
### Author Contributions

DM and AC designed the work and wrote the manuscript. SP and DM performed simulations and analyzed data. DM did the analytic blob theory calculations and also established the biological relevance of pur studies.

**Notes**: The authors declare no competing financial interest.

## ACKNOWLEDGEMENTS

A.C., with DST-SERB identification SQUID-1973- AC-4067, acknowledges funding by DST-India, project MTR/2019/000078 and discussions in meetings organized by ICTS, Bangalore, India.

# Supplementary: Tuning entropic repulsion between bacterial DNA-polymer loops achieves faster segregation and longitudinal organization of polymer segments under cylindrical confinement.

December 22, 2021

## 1   Details of segregation dynamics calculations

For the data presented in each of the plots of Fig.S1 we check if the distance between the center of masses is greater or equal to a certain threshold ($0.5L$). If that is the case we compute the average distance between the center of masses over the next $4 \times 10^7$ iterations. If the average is greater than the designated threshold then that particular case is deemed to be successful. From the pool of successful cases we extract the time of segregation. The fraction of 'successful' cases gives us the success rate of segregation.

We present the results of segregation times and success rates for all architectures in Fig.3 of the main manuscript. There the criterion for a particular case to be considered successful is: distance between center of masses of two overlapping polymers greater than or equal to $0.5L_o$.

Here we present data for a threshold of $0.4L_o$. Note that such a criterion implies that that we are looking at the timescales of partial segregation. We present cases both with and without the effect of Topoisomerase (chain crossing) incorporated.

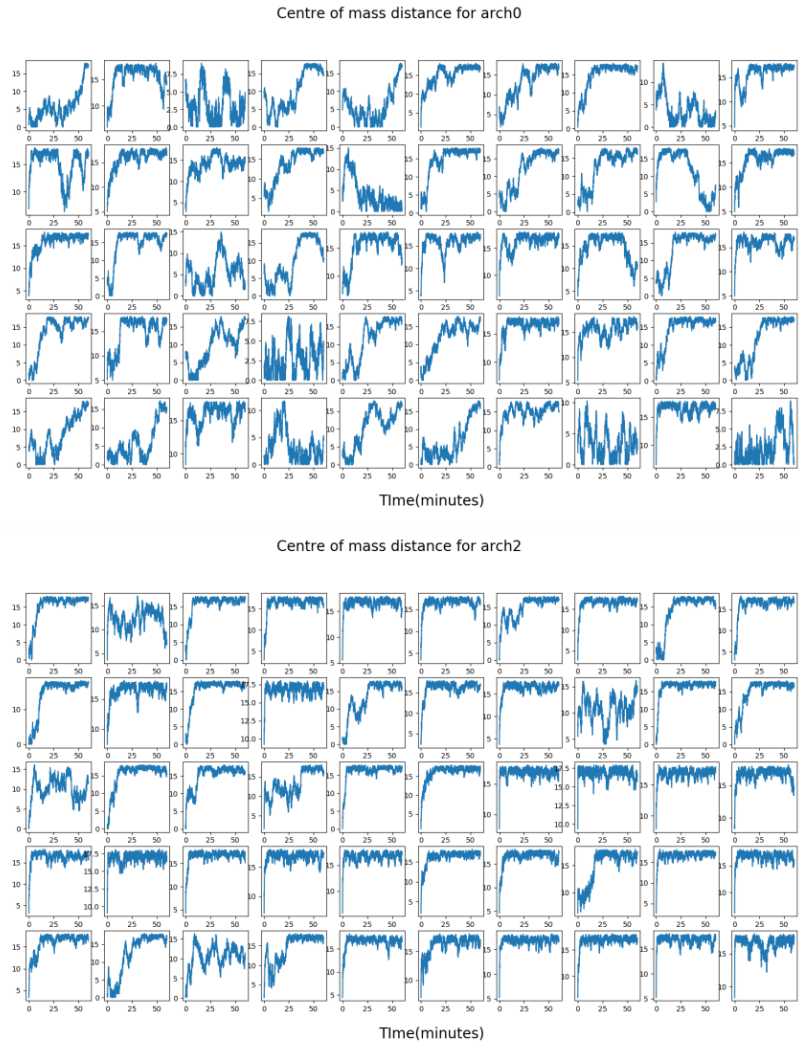

Figure S1: Plots (of 50 independent runs) for center of mass separation $D_{com}$ between two overlapping polymers as a function of time for (a) Arc-0 and (b)Arc-2 respectively. We emphasize that Arc-0 shows greater propensity to 'mix' after being 'demixed' as indicated by a large degree of fluctuations in $D_{com}$

## 2 *oriC* and *dif-ter* localisation for all architectures:without the effect of Topoisomerase (topological fluctuations) incorporated

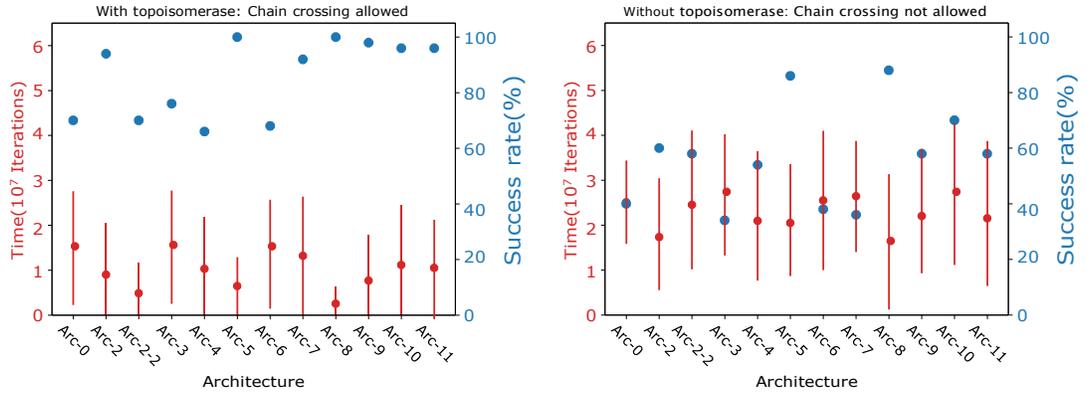

Figure S2: Fig. shows a comparison of segregation speeds for different architectures with (a) and without (b) the effect of Topoisomerase (topological fluctuations). The right y-axis shows the success rate of segregation which indicates what percentage of independent runs lead to segregation. The designated threshold criterion for considering two polymer chains to have spatially segregated is the mean (over $4 \times 10^7$ MCS) of the distance between the center of masses $D_{com}$ to have a value, greater or equal to $0.4L$, where $L$ denotes the length of the confining cylinder.

# 3   Localisation of loci: for all architectures (with the effect of Topoisomerase (topological fluctuations) incorporated)

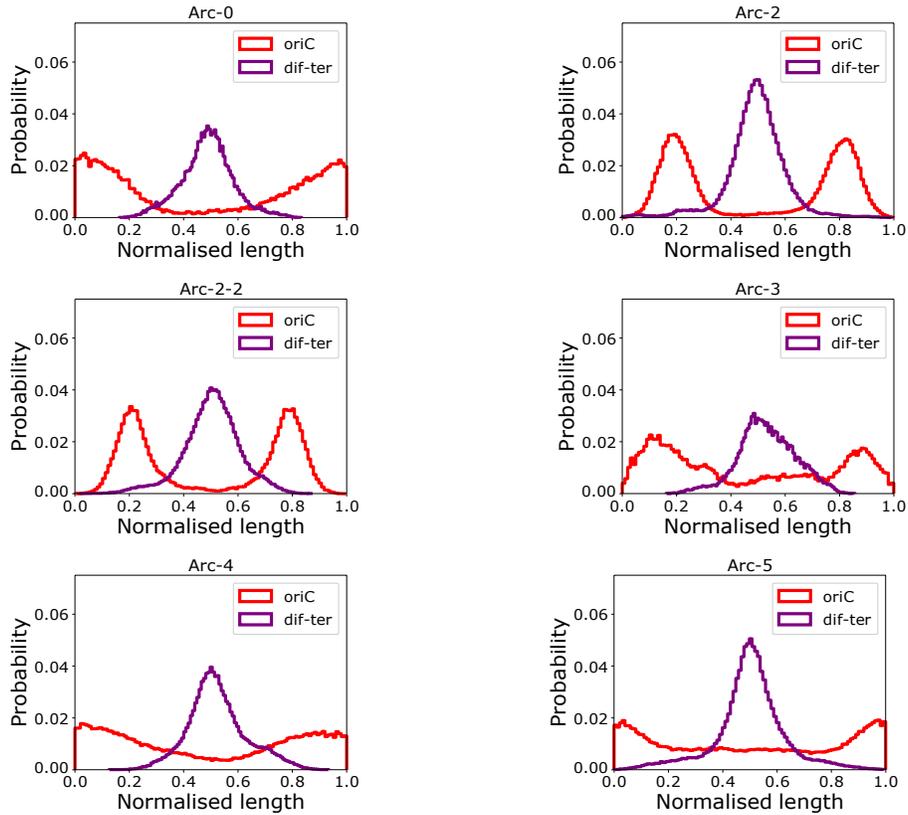

Figure S3: Fig. shows the distribution of the *oriC* positions for (a)Arc-0, (b)Arc-2, (c)Arc-3, (d)Arc-4 and (e)Arc-5. We note that arc2 shows a sharply peaked distribution of the *oriC* around the quarter positions. The data has been produced by averaging over the runs that successfully lead to segregation, from a pool of 50 independent runs.

We note that Arc-2 shows a sharply peaked distribution of the *oriC* position and the quarter positions and the dif locus shows a sharply peaked distribution around the mid-cell position. Note that we show the distribution of only one of the *dif* locus since the two *dif* loci are crosslinked and therefore are expected to have nearly overlapping distributions. The effect of Topoisomerase (topological fluctuations) has not been incorporated.

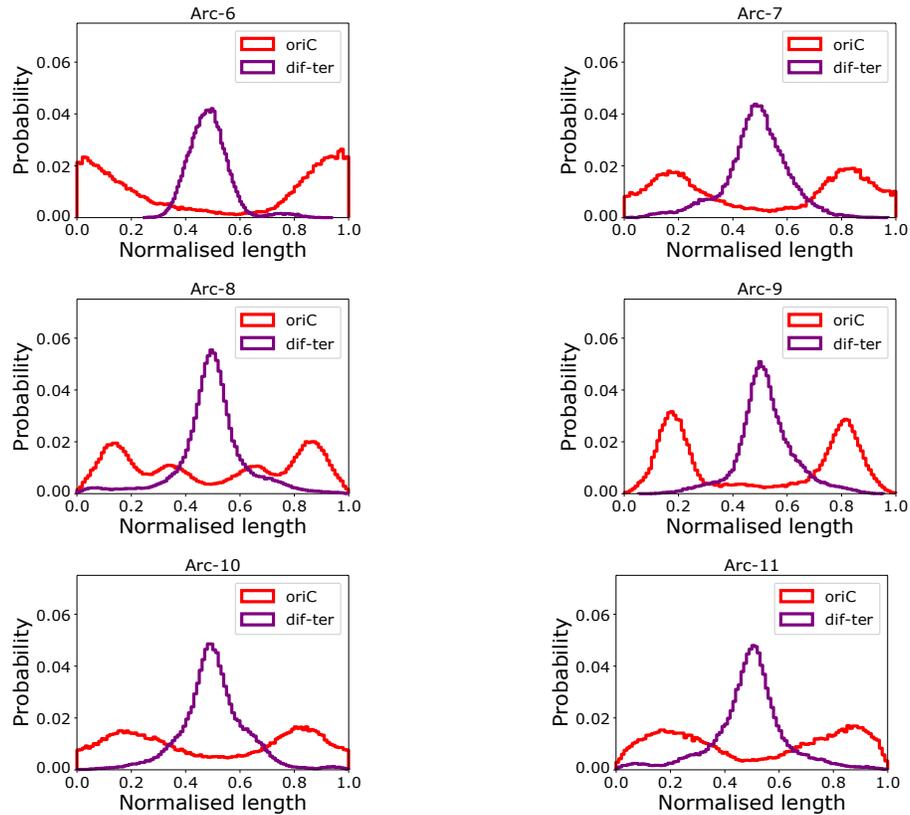

Figure S4: Fig.4 shows the distribution of the *oriC* positions for (a)Arc-6, (b)Arc-7, (c)Arc-8, (d)Arc-9, (e)Arc-10 and (f)Arc-11. We note that Arc-9 shows a sharply peaked distribution of the *oriC* position and the quarter posi- tions and the *dif* locus shows a sharply peaked distribution around the mid-cell position. Note that we show the distribution of only one of the *dif* locus since the two *dif* loci are crosslinked and therefore are expected to have nearly over- lapping distributions. The data has been produced by averaging over the runs that successfully lead to segregation, from a pool of 50 independent runs.

The effect of Topoisomerase (topological fluctuations) has not been incorpo- rated.

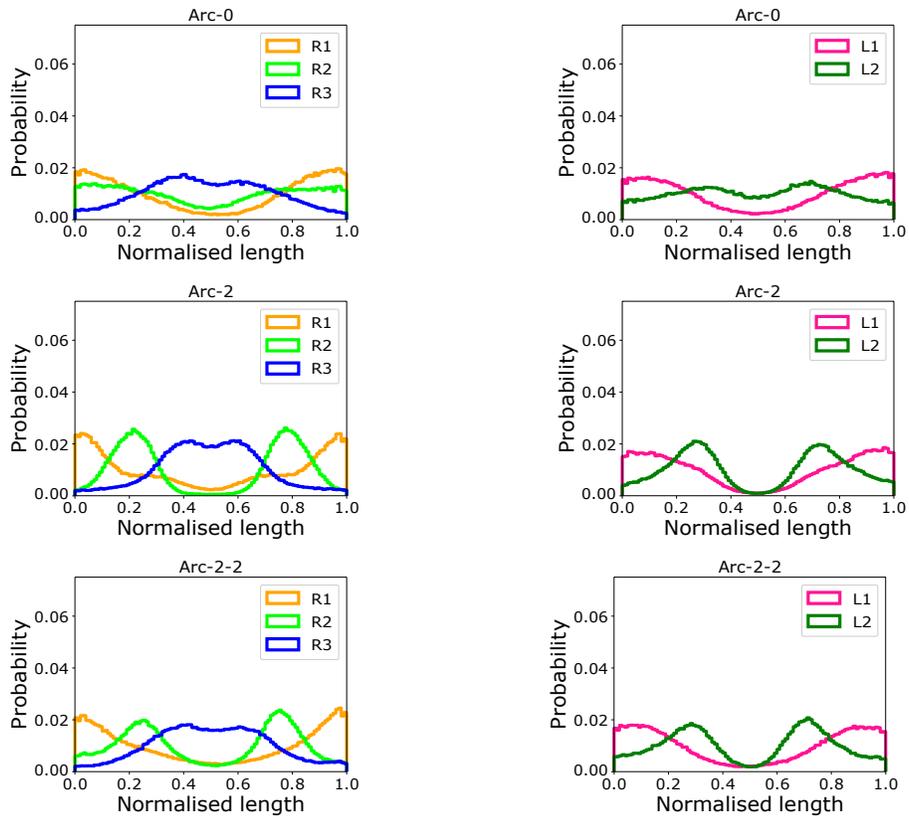

Figure S5: Fig shows the distribution of the loci positions for Arc-0, and (c),(d) show the distribution for Arc-2,. Subfigures (e) and (f) show loci distribution for Arc-2-2. The data has been produced by averaging over the runs that successfully lead to segregation, from a pool of 50 independent runs.
The effect of Topoisomerase (topological fluctuations) has been incorporated.

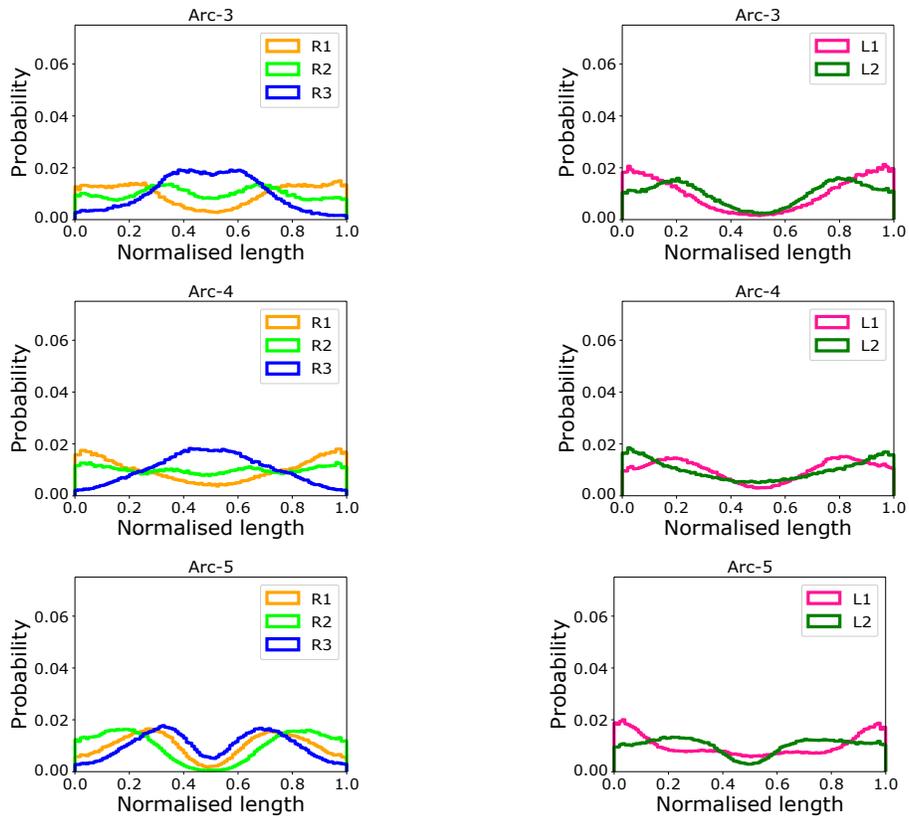

Figure S6: Fig. shows the distribution of the loci positions for Arc-3, and (c),(d) show the distribution for Arc-4. Subfigures (e) and (f) show loci distribution for Arc-5. The data has been produced by averaging over the runs that successfully lead to segregation, from a pool of 50 independent runs.
The effect of Topoisomerase (topological fluctuations) has been incorporated.

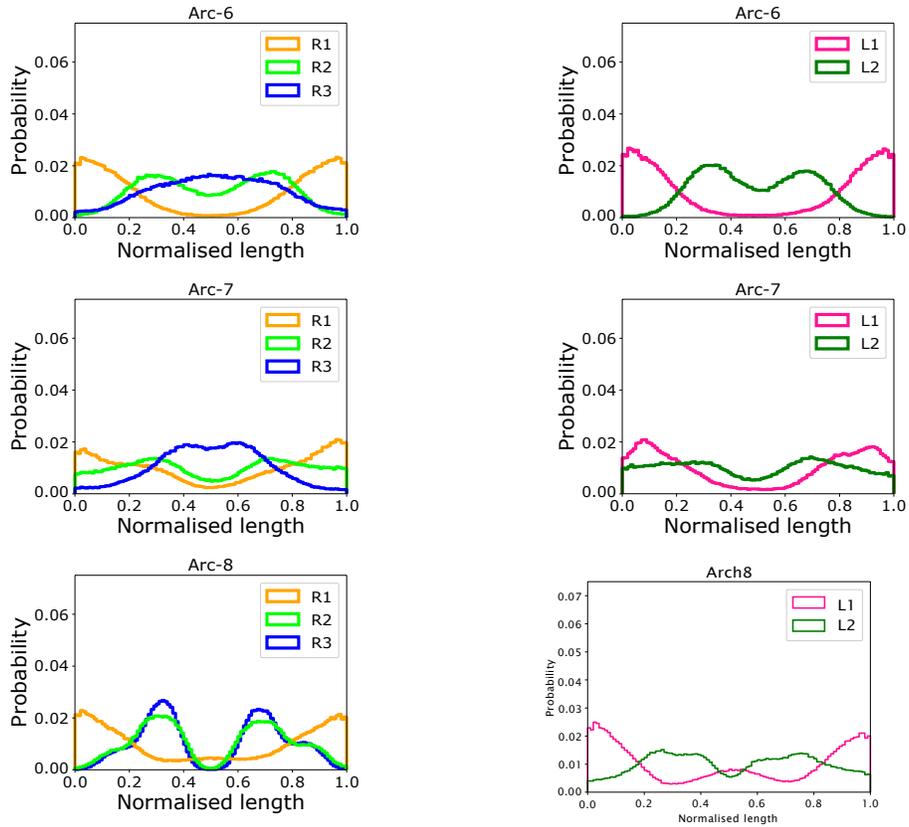

Figure S7: Fig. show the distribution of the loci positions for Arc-6, and (c),(d) show the distribution for Arc-7,. Subfigures (e) and (f) show loci distribution for Arc-8. The data has been produced by averaging over the runs that successfully lead to segregation, from a pool of 50 independent runs.
The effect of Topoisomerase (topological fluctuations) has been incorporated.

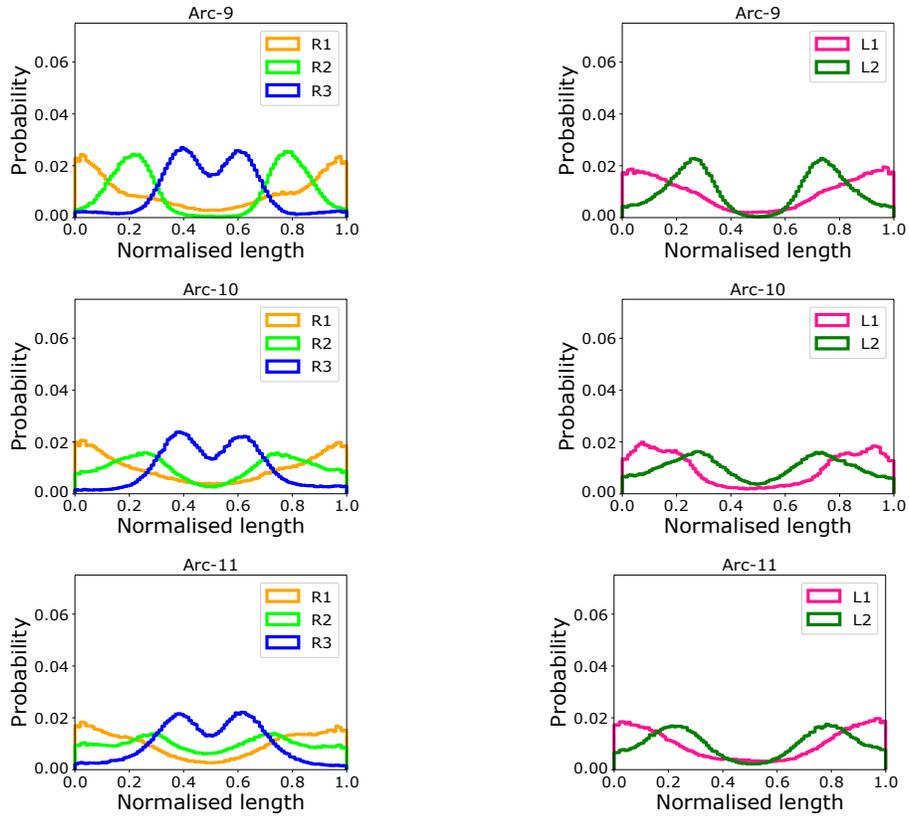

Figure S8: Fig. show the distribution of the loci positions for Arc-9, and (c),(d) show the distribution for Arc-10,. Subfigures (e) and (f) show loci distribution for Arc-11. The data has been produced by averaging over the runs that successfully lead to segregation, from a pool of 50 independent runs.
The effect of Topoisomerase (topological fluctuations) has been incorporated.

## 4 Localisation of loci: for all architectures (without the effect of Topoisomerase (topological fluctuations) incorporated)

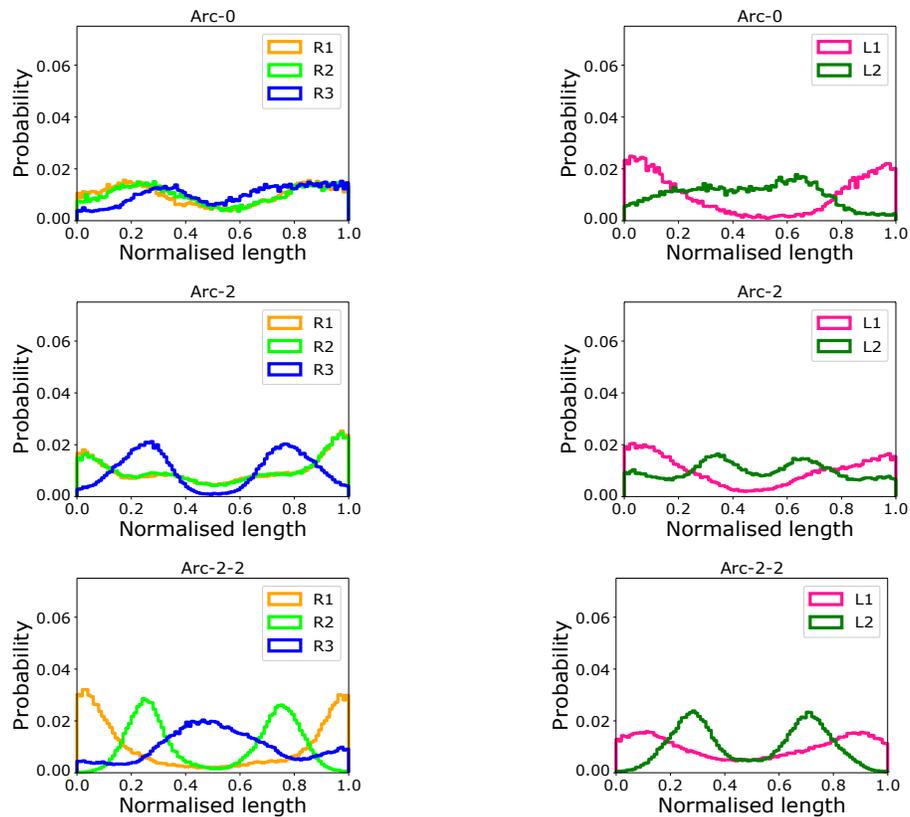

Figure S9: Fig. show the distribution of the loci positions for Arc-0, and (c),(d) show the distribution for Arc-2,. Subfigures (e) and (f) show loci distribution for Arc-2-2. The data has been produced by averaging over the runs that successfully lead to segregation, from a pool of 50 independent runs.
The effect of Topoisomerase (topological fluctuations) has not been incorporated.

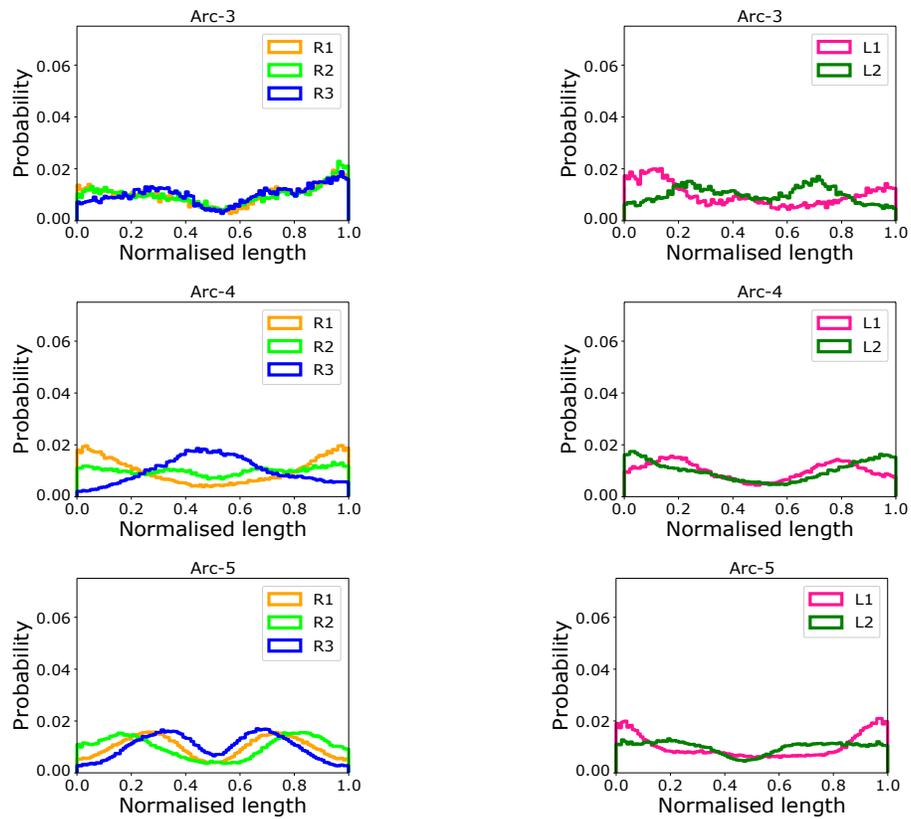

Figure S10: Fig. show the distribution of the loci positions for Arc-3, and (c),(d) show the distribution for Arc-4. Subfigures (e) and (f) show loci distribution for Arc-5. The data has been produced by averaging over the runs that successfully lead to segregation, from a pool of 50 independent runs.

The effect of Topoisomerase (topological fluctuations) has not been incorporated.

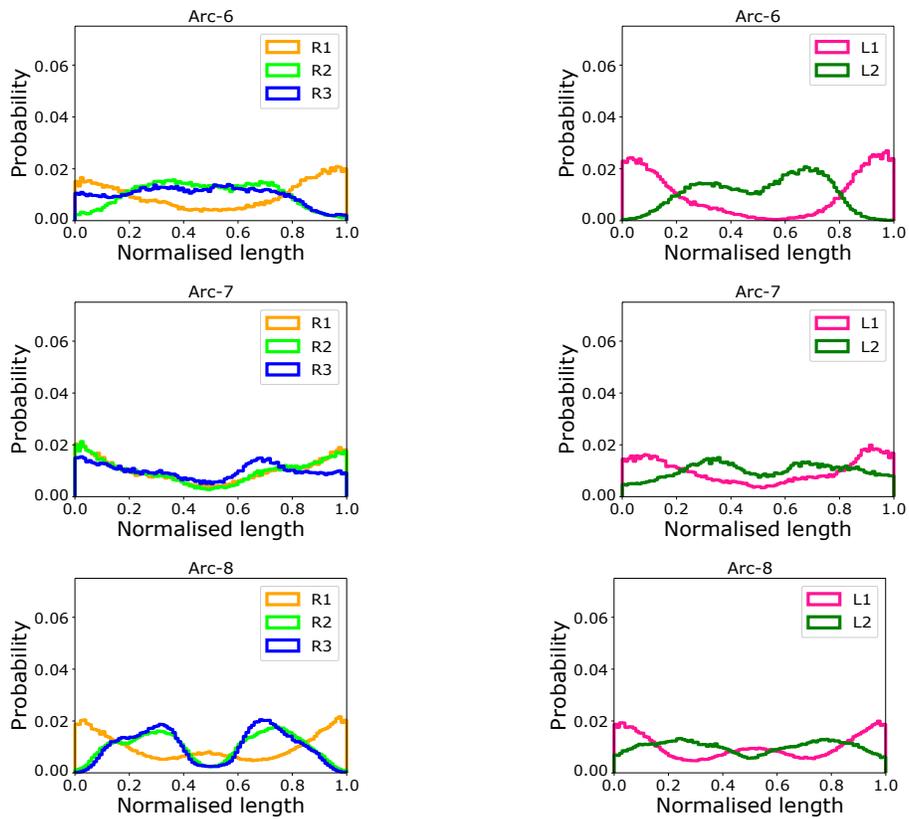

Figure S11: Fig. shows the distribution of the loci positions for Arc-6, and (c),(d) show the distribution for Arc-7,. Subfigures (e) and (f) show loci distribution for Arc-8. The data has been produced by averaging over the runs that successfully lead to segregation, from a pool of 50 independent runs.

The effect of Topoisomerase (topological fluctuations) has not been incorporated.

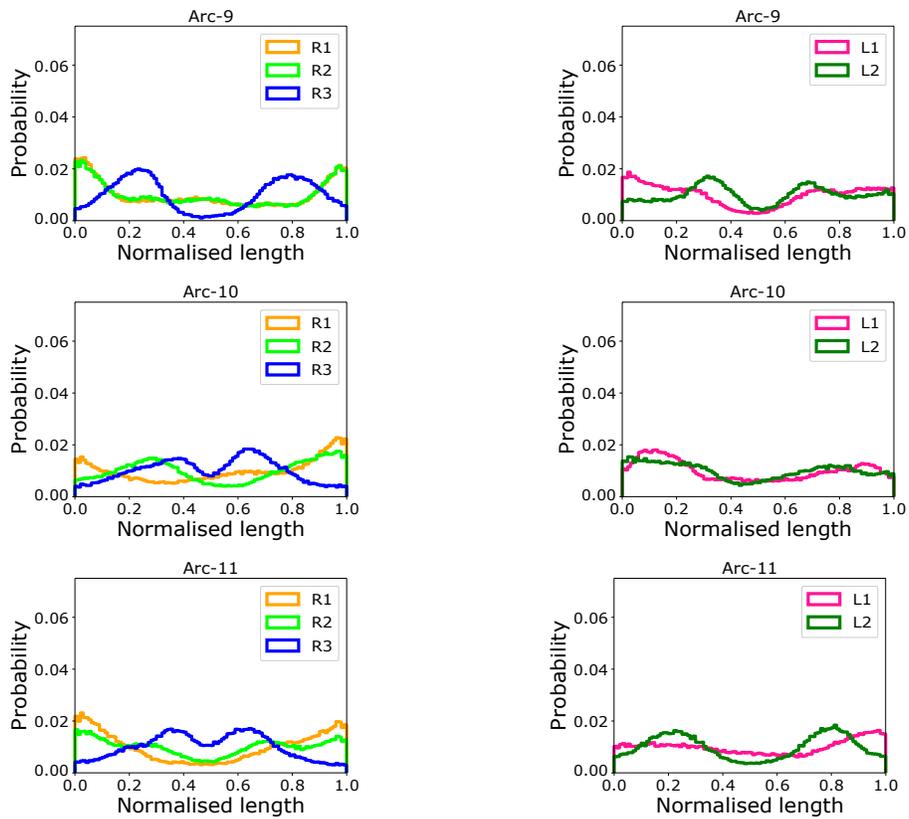

Figure S12: Fig. shows the distribution of the loci positions for Arc-9, and (c),(d) show the distribution for Arc-10,. Subfigures (e) and (f) show loci distribution for Arc-11. The data has been produced by averaging over the runs that successfully lead to segregation, from a pool of 50 independent runs. The effect of Topoisomerase (topological fluctuations) has not been incorporated.

# 5 Details of free energy calculations

## 5.1 Free energy difference between the overlapped and the segregated states for Arc-0 polymer architecture confined in a cylinder of diameter D.

### i. Segregated state

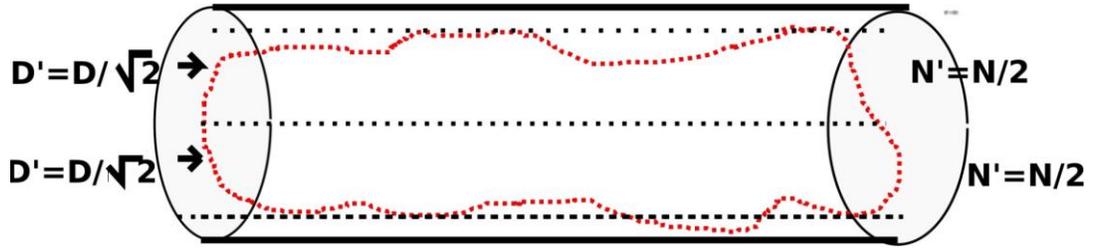

Figure S13: Fig. shows a schematic representation of single ring polymer with $N = 500$ monomers. Each arm of the ring polymer from monomer number 1 to 250 with $N' = N/2 = 250$ monomers can be visualized as a linear segment constrained in an effective tube of diameter $D' \approx D/\sqrt{2}$.

To calculate the Helmholtz free energy $F_{seg}$ for two segregated ring polymers, conjoined at the ter and confined in a cylinder of diameter $D$, we use the "renormalized flory approach". We calculate the free energy corresponding to just one confined ring polymer given by $F_{conf}$. The free energy of the two segregated ring polymers is given by

$$F_{seg} = 2F_{conf}$$

To calculate $F_{conf}$, we first assume that each linear segment containing 'N/2' monomers is constrained to a tube of an effective diameter of $D' \approx D/\sqrt{2}$ [?]. A segment (blob) of a self avoiding polymer chain spans diameter $D'$ with $a \times g^{3/5}$ monomers. Thus the total number of monomers in each 'blob' is given by

$$g = (D/\sqrt{2})^{5/3}. \tag{1}$$

and the total number of self-avoiding blobs within a cylinder, for two linear chain of $N'$ monomers is therefore given by ,

$$n_{bl} = 2 \times (N'/g) = 2 \times \frac{N}{2} \times (\sqrt{2}/D)^{5/3}) \tag{2}$$

If the free energy associated with each blob is given by $f_{bl}$ (later taken $\approx k_B T$), then the free energy of the segregated state $F_{seg}$ with 2 polymers is given by

$$F_{seg} = 2F_{conf} = 2 \times \frac{N}{D^{5/3}} \times 2^{5/6} \times f_{bl} \tag{3}$$

## ii. Free energy of 2 rings polymer (Arc-0 architecture) in overlapped state confined in a cylinder of diameter $D$.

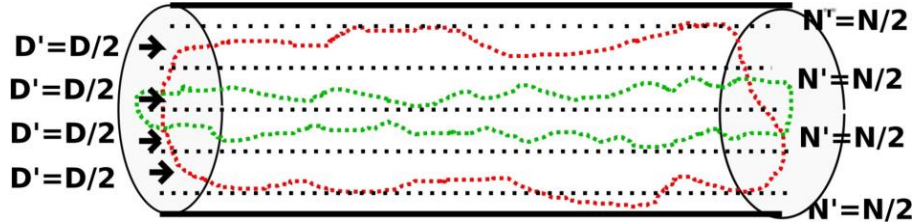

Figure S14: Fig. shows a schematic representation of two overlapping ring polymers (shown in red and green) where each arm can be visualized as a linear segment constrained in an effective tube of diameter $D^{'} \approx D/2$

To calculate the Helmholtz free energy $F_{ov}$ for two overlapping ring polymers (conjoined at the ter) we assume that each arm of a single ring polymer containing $N^{'} = N/2 = 250$ monomers occupies an effective tube of diameter $D^{'} = D/2$. Note that there are 4 such linear arms since there are two overlapping ring polymers. The cross-sectional area of the four tubes (each containing a polymer arm) remains $4 \times \pi(D^{'})^2 = \pi D^2$. Thus the total number of monomers in each 'blob' is given by

$$g = (D/2)^{5/3}. \tag{4}$$

and the total number of blobs for four polymer-arms are therefore given by ,

$$n_{bl} = 4 \times (N^{'}/g) = 4 \times \frac{N}{2} \times (2/D)^{5/3} \tag{5}$$

If the free energy associated with each blob is given by $f_{bl}$, then the free energy of the overlapped state $F_{ov}$ is given by

$$F_{ov} = 2N \times (2/D)^{5/3} \times f_{bl} \tag{6}$$

Thus the **Free energy difference per monomer**, between the overlapped and the segregated state for two Arc-0 (ring polymers) in a cylinder of diameter $D$, is given by

$$\delta F/N = (F_{ov} - F_{seg})/N = 2.78 f_{bl}/D^{5/3} \tag{7}$$

## 5.2 Free energy difference between the overlapped and the segregated states of two Arc-2 polymers confined within a cylinder of diamter $D$.

A polymer with the Arc-2 architecture has two loops of 125 monomers each in a chain of 500 monomers. Thus there are $N/4$ monomers in each loop of a polymer having Arc-2 architecture, while the bigger loop has $N/2$ monomers, refer Fig.2 of the main manuscript.

### i. Segregated state

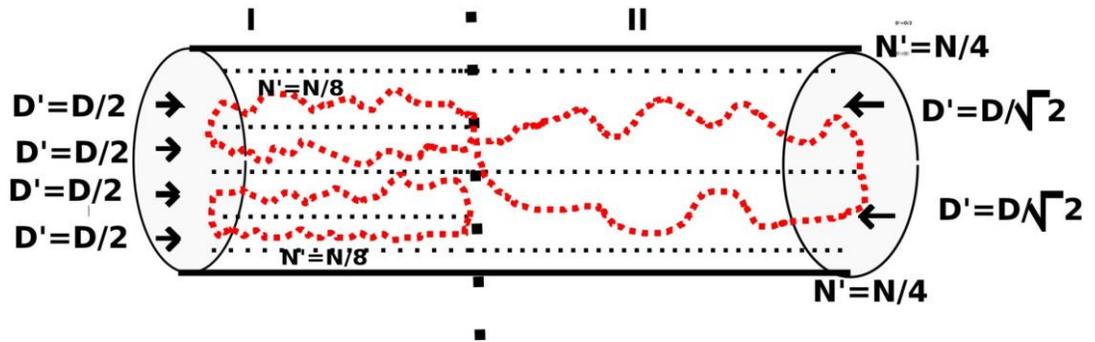

Figure S15: Fig. shows a schematic representation of single polymer in the Arc-2 configuration. The cylinder can be divided into two sections: section-I containing the two smaller loops and Section-II containing the larger loop with $N/2$ monomers. The confined larger loop (which itself is a ring polymer with $N/2$ monomers) can be visualized as two linear segments with $N/4$ monomers in each chain. Each linear segment is confined in a effective tube of diameter $D^{'} \approx D/2$. In region I, there are the two smaller loops (ring polymers) with $N/4$ monomers each. Each loop can be thought of as containing two linear segments with $N/8$ monomers, each occupying effective tubes of diameter $D^{'} \approx D/2$.

. In order to calculate the Helmholtz free energy for two segregated polymers (conjoined at the ter) of architecture Arc-2 using the "renormalized flory approach", we need to calculate the free energy corresponding to just one confined polymer given by $F_{conf}$ . The free energy of the two segregated ring polymers is given by $F_{seg} = 2F_{conf}$.

· To calculate $F_{conf}$, we assume that $F_{conf}$ is given by the sum of free energies in section-I and section-II: $F_{conf} = F_I + F_{II}$, where $F_I$ and $F_{II}$ denote the contributions to the free energy from sections I and II, as shown in Fig. S15.

· In section-I, each linear segment of $N/8$ monomers is constrained in a tube of an effective diameter of $D^{'} \approx D/2$. Thus the number of monomers in each

'blob' is given by,

$$g = (D/2)^{5/3}.$$ (8)

and the total number of blobs from 4 linear chains with $N/8$ monomers is given by ,

$$n_{bl} = 4 \times \frac{N}{8} \times (2/D)^{5/3}$$ (9)

If the free energy associated with each blob is given by $f_{bl}$, then the free energy $F_I$ is given by

$$F_I = (N/2)/D^{5/3} \times 2^{5/3} f_{bl}$$ (10)

· In region II, each linear segment of $N/4$ monomers is constrained in a tube of an effective diameter of $D' \approx D/\sqrt{2}$. The number of monomers in each 'blob' is:

$$g = (D/\sqrt{2})^{5/3}.$$ (11)

and the total number of blobs from two chains each with ($N/4$) monomers is:

$$n_{bl} = 2 \times \frac{N}{4} \times (\sqrt{2}/D)^{5/3}$$ (12)

If the free energy associated with each blob is given by $f_{bl}$, then $F_{II} = n_{bl}f_{bl}$:

$$F_{II} = 2 \times \frac{N}{2} \times (\sqrt{2}/D)^{5/3} \times f_{bl}$$ (13)

Thus for 2 segregated Arc-2 polymers which are next to each other in a cylinder of diameter $D$,

$$F_{seg} = 2F_{conf} = 2(F_I + F_{II})$$ (14)

. *Other possible configurations:* We have assumed a particular configuration of the 'Arc-2' polymer where the two smaller loops occupy section-I and the bigger loop occupy section-II of cylinder. To rule out the possibility that one smaller loop occupies section-I, and the other smaller loop overlaps with the bigger loop in section-II of cylinder, we have calculated (using the method outlined above) that the free energy for confinement for such a configuration is higher than the configuration discussed in detail above. Other possible configurations, such as that of all loops occupying the same cylinder half, also have higher values of free energy. For one small loop occupying the same half as the bigger loop, the free energy of confinement ($F_{conf}$) is increased by $0.3N/D^{5/3}f_{bl}$.

## ii. Overlapped state

In order to calculate the Helmholtz free energy using the "renormalized flory approach", for two overlapped polymers (conjoined at the ter) of architecture Arc-2, we once again employ the same approach as before. Thus, To calculate $F_{ov}$ we first assume that $F_{ov}$ is given by, $F_{ov} = F_I + F_{II}$ , where $F_I$ and $F_{II}$ denote the contributions to the free energy from regions I and II, as shown in Fig.**??**.

In region I each linear segment of $N/8$ monomers is constrained in a tube of an effective diameter of $D^{'} \approx D/2\sqrt{2}$. Thus the number of monomers in each 'blob' is given by,

$$g = (D/2\sqrt{2})^{5/3}. \tag{15}$$

and the total number of blobs are therefore given by ,

$$n_{bl} = 8 \times N/8 \times (2\sqrt{2})^{5/3}/D^{5/3} \tag{16}$$

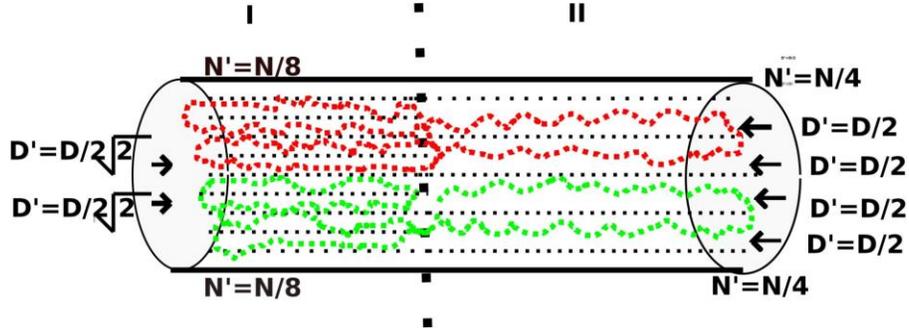

Figure S16: Fig. shows a schematic representation of two overlapping ring polymers (shown in red and green) where each arm can be visualized as a linear segment constrained in an effective tube of diameter $D^{'} \approx D/2$

If the free energy associated with each blob is given by $f_{bl}$, then the free energy $F_I$ is given by

$$F_I = N \times (2\sqrt{2})^{5/3}/D^{5/3}f_{bl} \tag{17}$$

In region II each linear segment of $N/4$ monomers is constrained in a tube of an effective diameter of $D^{'} \approx D/2$. Thus the number of monomers in each 'blob' is given by,

$$g = (D/2)^{5/3}. \tag{18}$$

and the total number of blobs are therefore given by ,

$$n_{bl} = 4 \times N/4 \times 2^{5/3}/D^{5/3} \tag{19}$$

If the free energy associated with each blob is given by $f_{bl}$, then the free energy $F_{II}$ is given by

$$F_{II} = N \times 2^{5/3}/D^{5/3}f_{bl} \tag{20}$$

Thus,

$$F_{ov} = F_I + F_{II} \tag{21}$$

Thus the free energy difference between the overlapped and the segregated state, per monomer, corresponding to Arc-2 , is given by

$$\delta F/N = (F_{ov} - F_{seg})/N = 3.87 f_{bl}/D^{5/3} \tag{22}$$

Thus in this way we can calculate $\delta F/N$ can be calculated for each polymer architecture using the "renormalized flory approach" where each loop is treated as linear chains trapped in an effective tube.

## 5.3 $\delta F/N$ corresponding to all polymer architectures

In the following table we present the analytically calculated $\delta F/N$ for different architectures, using the methods outlined in the previous sections.

| Architecture | $\delta F/N$ |
|---|---|
| Arc-0 | $2.78 f_{bl}/D^{5/3}$ |
| Arc-2 | $3.87 f_{bl}/D^{5/3}$ |
| Arc-2-2 | $4.0 f_{bl}/D^{5/3}$ |
| Arc-3 | $2.07 f_{bl}/D^{5/3}$ |
| Arc-4 | $4.90 f_{bl}/D^{5/3}$ |
| Arc-6 | $3.30 f_{bl}/D^{5/3}$ |
| Arc-7 | $3.30 f_{bl}/D^{5/3}$ |
| Arc-8 | $5.30 f_{bl}/D^{5/3}$ |
| Arc-9 | $3.84 f_{bl}/D^{5/3}$ |
| Arc-10 | $3.23 f_{bl}/D^{5/3}$ |
| Arc-11 | $2.85 f_{bl}/D^{5/3}$ |

The above table indicates the $\delta F/N$ corresponding to each polymer architecture, except Arc-5. This is because Arc-5 has a 'nested loop' like architecture which cannot be treated by the simplistic "renormalized flory approach".

In Fig.7 of the main manuscript we present $\delta F/N$ for each polymer architecture for an arbitrary set of parameters: $D = 10a$ and $f_{bl} = k_B T$.